\documentclass[11pt]{article}

\usepackage{amsfonts,amsmath,amssymb,graphics,bm,amsthm,xcolor}

\theoremstyle{definition}
\newtheorem{example}{Example}

\newtheorem{defn}{Definition}

\newtheorem{thm}{Theorem}
\newtheorem{result}{Result}

\newcommand{\gapn}{\vspace{3mm}\noindent}

\newcommand{\I}{\mathrm{i}}
\newcommand{\pr}{\mathbf{P}}
\newcommand{\ex}{\mathbf{E}}

\newcommand{\wex}{\widehat{\mathbf{E}}}
\newcommand{\wpr}{\widehat{\mathbf{P}}}

\newcommand{\notthis}[1]{}

\newcommand{\indic}[1]{\mathbf{1}[#1]}
\newcommand{\half}{\frac{1}{2}}

\newcommand{\shalf}{{\textstyle\frac{1}{2}}}

\newcommand{\pderiv}[2]{\frac{\partial{#1}}{\partial{#2}}}

\newcommand{\hsigma}{\widetilde{\sigma}}

\newcommand{\Xup}{X_\textrm{up}}
\newcommand{\Xdn}{X_\textrm{dn}}
\newcommand{\Xfwd}{X_\textrm{fwd}}
\newcommand{\Yup}{Y_\textrm{up}}
\newcommand{\Ydn}{Y_\textrm{dn}}

\newcommand{\Vup}{V_\textrm{up}}
\newcommand{\Vdn}{V_\textrm{dn}}
\newcommand{\Wup}{W_\textrm{up}}
\newcommand{\Wdn}{W_\textrm{dn}}
\newcommand{\phat}{\widehat{p}\,}

\newcommand{\mma}{\mathfrak{B}}

\begin{document}

\title{\bf Black--Scholes without stochastics or PDEs}
\author{Richard J. Martin\footnote{Department of Mathematics, Imperial College London, South Kensington, London SW7 2AZ, UK} }
\maketitle

\begin{abstract}

We show how to derive the Black--Scholes pricing formula and its generalisation to the `exchange-option' (to exchange one asset for another) via the continuum limit of the Binomial tree. No knowledge of stochastic calculus or partial differential equations is assumed, as we do not use them.

\end{abstract}

A standard result of option pricing theory, taught in all quantitative finance courses, is the Black--Scholes formula, which gives the no-arbitrage price of a European call or put option on a non-dividend-paying stock, subject to a variety of simplifying assumption (constant volatility geometric Brownian motion etc). While the mathematical preliminaries for this are not especially demanding, we think it worth pointing out that It\^o calculus and diffusive PDEs are not strictly necessary. It can in fact be done via the continuum limit of a binomial tree, with some algebra and calculus. For one thing, such a development keeps an important point in focus: we need the underlying to be continuously tradable.

However, there is another benefit, in that for some applications, particularly in fixed income, the option is to exchange one asset for another ($X,Y$ say). A common route to pricing this contingent claim is to start with the Black--Scholes analysis and then use a change of num\'eraire to write down the dynamics of $X_t/Y_t$. The original paper is by Margrabe \cite{Margrabe78}; see also \cite[\S19.3]{Bjork98}.
But this is a rather roundabout route and the resulting formula (\ref{eq:ch5.bs0}) looks so neat that one is led to wonder whether it could have been derived directly---to which the answer is yes, it can.

There are so many general introductions to derivative pricing that it is unnecessary for us to repeat background information: we refer the reader to, for example, \cite{Bjork98,Hull_OFOD}. The route we take is: one-period model and martingale probabilities; binomial model; two-asset binomial model; continuum limit. We thereby end up with the exchange-option result directly.

\section{One-period models and their continuum limits}

\subsection{One-period (binary) model}

Consider in a one-period model (time $[0,\tau]$) with two future states, $u$ and $d$, the pricing of a claim worth $\Vup$ or $\Vdn$ according as the market moves up or down:

\begin{center}
\begin{tabular}{rrr}
State & $X$ & Payout \\
\hline
$u$ & $\Xup$ & $\Vup$ \\ 
$d$ & $\Xdn$ & $\Vdn$ 
\end{tabular}
\end{center}

\noindent
Let us consider the replication of this claim using a riskfree asset and a forward contract on $X$. By writing
\begin{eqnarray*}
\Vup &=& a_X \Xup + a_\textrm{rf} \\
\Vdn &=& a_X \Xdn + a_\textrm{rf}
\end{eqnarray*}
we can solve for the amounts $a_X$ and $a_\textrm{rf}$ to obtain:
\[
a_X = \frac{\Vdn-\Vup}{\Xdn-\Xup}, \qquad a_\textrm{rf} = \frac{\Vdn \Xup - \Vup \Xdn}{\Xup - \Xdn} .
\]
 The value of the claim today must be the same as that of the replicating portfolio, which is
\[
\big(a_X \Xfwd + a_\textrm{rf}\big) B(\tau)
\]
which can be written as
\begin{equation}
\boxed{V_0 = \big( \phat \Vup + (1-\phat) \Vdn \big) B(\tau)}
\label{eq:ch5.repl}
\end{equation}
with
\begin{equation}
\boxed{\phat = \frac{ \Xfwd - \Xdn }{ \Xup - \Xdn }, \qquad
1-\phat = \frac{ \Xup - \Xfwd }{ \Xup - \Xdn }}.
\label{eq:ch5.probs}
\end{equation}
By using a forward contract on $X$, we have covered the case when $X$ pays a coupon or dividend.

Equation (\ref{eq:ch5.repl}) has the form of a discounted expected payoff, provided our understanding of the term `expectation' does not mean anything to do with `what we expect' or any kind of subjective judgement---rather it is a weighing mechanism using weights $\phat$, $1-\phat$.
These are known as the \emph{martingale probabilities} (of moving up or down). They have the property that if the claim is just equal to the underlying asset, so that it is simply a forward on the that asset, then it is priced correctly:
\begin{equation}
\phat \Xup + (1-\phat) \Xdn = \Xfwd.
\label{eq:ch5.mtg}
\end{equation}

\begin{defn}
We say that the forward price of $X$ is a \emph{martingale under $\wpr$}. This means that the expectation of its future value RHS(\ref{eq:ch5.mtg}) is its current value LHS(\ref{eq:ch5.mtg}), providing the probabilities $\phat$, $1-\phat$ are used in calculating the expectation.
\end{defn}

We reemphasise that the martingale probabilities have nothing to do with the real-world probabilities and in effect the martingale condition is simply the calibration of a weighing-machine, so that when one places a tradable asset on the scales, one obtains today's value (discounting aside). Indeed, the martingale probabilities have a `self-righting effect' in the following sense. Suppose that I increase $\Xup$ and $\Xdn$, for simplicity by the same amount. Clearly $\phat$ goes down and $1-\phat$ goes up. If this did not happen, the model would disagree on the forward price of the asset. Rather like tipping a beaker of water, the liquid finds its own level. That said, water will be spilt if $\Xup,\Xdn$ are moved up or down too far, pushing the martingale probabilities outside the range $[0,1]$: if $\phat>1$ then there is a riskfree profit from buying the asset, and if $<0$ there is a riskfree profit from shorting it. Therefore, absence of arbitrage requires the martingale probabilities to be \emph{equivalent} to the real-world ones, which means that they must agree on what is possible and what is not (they don't have to agree on how likely anything is, though).

\begin{example}
Let the forward price of $X$ be 101 and the up/down values be 104 or 99 respectively, and let the discount-factor be $\frac{100}{101}$. Let the claim pay out 3 (up) or 2 (down). The martingale probabilities are 0.4 (up), 0.6 (down), the delta (amount of risky asset needed for replication) is $a_X=0.2$, and the claim today is worth 2.38.
$\Box$
\end{example}

\subsection{From binary to binomial}

If we take a one-step model and chain many steps together we get a \emph{tree}. In a geometric model, which is the usual mode of development, making the steps work as follows: an up-move causes a multiplication by a specified amount $\mathfrak{u}>1$, and down-move causes a multiplication by another specified amount $\mathfrak{d}<1$. Importantly if we move up and then down, or down and then up, we end up with the same result ($\times \mathfrak{ud}$ in each case). The tree is then said to be \emph{recombining}, so that after $n$ steps there are only $n+1$ possible terminal values rather than $2^n$. We now dispense with the notation $\Xup$, $\Xdn$, as we want to use subscripts to denote time.
Provided we use the same up/down probabilities everywhere, the resulting probability distribution is Binomial: 
\[
\pr\big(X_n = X_0 \mathfrak{u}^j \mathfrak{d}^{n-j}\big)  = {n \choose j} p^j (1-p)^{n-j}, \qquad 0\le j \le n
\]
where ${n \choose j}$ denotes as usual the Binomial coefficient $\frac{n!}{(n-j)!j!}$.
For this reason the model is called the \emph{binomial tree}. 

If using martingale probabilities, we replace $\pr$ with $\wpr$ and $p$ with $\phat$ and everything works providing the martingale probabilities are the same everywhere, which for constant interest rates they will be.

Contingent claims can then be valued by either forward or backward induction. Forward, and one calculates the probabilities through the tree and then evaluates the expected payoff at the end. Backward, and one calculates the payoff in each state at the end, and then works backwards to the beginning.
\[
\boxed{\mbox{Probabilities propagate forwards. Expectations propagate backwards.}}
\]
The inductions are known as the Chapman-Kolmogorov equations.

\begin{example}\label{ex:ch6.tree}
Suppose $X_0=100$, and set up a 6-step tree with discount-factor 0.996 per step. Value the contingent claim $\varphi(X_T)=\min(X_T,101)$, where $T$ denotes step \#6, using a geometric tree with $\mathfrak{u}=1.02$, $\mathfrak{d}=0.98$.
\end{example}

\noindent
Solution. The up/down probabilities are $0.6004,0.3996$ at all points on the tree. The following tableau gives the values of $X_n$ (time going acrossways) and payoff $V$ at the last step, together with the probability of arriving at each point at time step 6. In other words, the forward-induction method.

\begin{center}\small
\begin{tabular}{rrrrrrrrr}
\hline
0	&1	&2	&3	&4	&5	&6	&Payoff &	Prob. \\
\hline
100.00	&102.00	&104.04	&106.12	&108.24	&110.41	&112.62	&101.00	&0.0468\\
&	98.00	&99.96	&101.96	&104.00	&106.08	&108.20	&101.00	&0.1871\\
&&		96.04	&97.96	&99.92	&101.92	&103.96	&101.00	&0.3112\\
&&&			94.12	&96.00	&97.92	&99.88	&99.88	&0.2762\\
&&&&				92.24	&94.08	&95.96	&95.96	&0.1379\\
&&&&&					90.39	&92.20	&92.20	&0.0367\\
&&&&&&						88.58	&88.58	&0.0041\\ \hline
\end{tabular}\end{center}

\noindent
The expectation of the payoff is 99.62, and discounting by $(0.996)^6$ gives 97.26, so this is the value we seek. 

Now for backwards. In this tableau, the terminal payoff values (time step 6) is shown as $\varphi(X_6)$. We then roll the calculation of $V$ back through the tree until we get to time zero: 

\begin{center}\small
\begin{tabular}{rrrrrrr}
\hline
0	&1	&2	&3	&4	&5	&6	\\
\hline
\textcolor{red}{97.26}	& 98.28	& 99.10	& 99.72	& 100.19	& 100.60	& 101.00\\
&	96.70	& 98.03	& 99.16	& 100.02	& 100.60	& 101.00\\
&&		95.66	& 97.33	& 98.86	& 100.15	& 101.00\\
&&&			94.12	& 96.00	& 97.92	& 99.88\\
&&&&			92.24	& 94.08	& 95.96\\
&&&&&					90.39	& 92.20\\
&&&&&&						88.58\\
 \hline
\end{tabular}\end{center}

\noindent
This gives the same result, 97.26. $\Box$

\gapn

\subsection{Rolled-up money market account}

In general the interest rate will vary stochastically.
To represent this discounting effect we divide by the \emph{rolled-up money-market account} defined as follows:
\[
\mma_T = (1+r_1\tau)(1+r_2\tau) \cdots (1+r_n\tau), \qquad \tau=T/n.
\]
Discounting from time $T$ today is achieved simply by dividing through by $\mma_T$.
The expectation of some contingent claim is therefore written
\begin{equation}
V_0 = \wex_0 [V_T/\mma_T].
\end{equation}


\subsection{Replication theory for two assets (homothetic payoff)}

Let us rework the replication argument from before, but instead deal with a pair of risky assets $X,Y$. The claim depends on their values in a particular way: loosely, it has to be `proportional' in the sense that if $X$ and $Y$ are both scaled up by some constant $\lambda$ then $V$ scales by the same amount\footnote{This is known as a \emph{homothetic} or \emph{1-homogeneous} function.}. This is equivalent to
\[
V = Y\times \textrm{func}\left(\frac{X}{Y}\right) .
\]
(Some condition is necessary, as otherwise we are attempting to price an arbitrary claim on two risky assets using only a one-dimensional argument.)
Then
\begin{eqnarray*}
\Vup &=& a_X \Xup + a_Y \Yup \\
\Vdn &=& a_X \Xdn + a_Y \Ydn
\end{eqnarray*}
so that
\[
a_X = \frac{\Vup\Ydn-\Vdn\Yup}{\Xup\Ydn-\Xdn\Yup}, \qquad 
a_Y = \frac{\Vdn\Xup-\Vup\Xdn}{\Xup\Ydn-\Xdn\Yup}
\]
and the value of the claim today is
\[
V_0 =
\frac{ \displaystyle\frac{\Vup}{\Yup} \left( X_0 - Y_0 \frac{\Xdn}{\Ydn} \right) }{ \displaystyle \frac{\Xup}{\Yup} - \frac{\Xdn}{\Ydn} }
+
\frac{ \displaystyle \frac{\Vdn}{\Ydn} \left( \frac{\Xup}{\Yup} Y_0 - X_0 \right) }{ \displaystyle \frac{\Xup}{\Yup} - \frac{\Xdn}{\Ydn} }
\]
Writing $W=X/Y$, we have
\[
\frac{V_0}{Y_0} = 
\left(\frac{V}{Y}\right)_\textrm{up} \frac{W_0-\Wdn}{\Wup-\Wdn} + 
 \left(\frac{V}{Y}\right)_\textrm{dn} \frac{\Wup-W_0}{\Wup-\Wdn} 
.
\]
We can therefore write this as a discounted expectation, identifying martingale probabilities as:
\begin{equation}
\phat_Y = \frac{W_0-\Wdn}{\Wup-\Wdn}, \qquad
1-\phat_Y = \frac{\Wup-W_0}{\Wup-\Wdn}.
\label{eq:ch5.probs2}
\end{equation}

\noindent
Note carefully all of the following:
\begin{itemize}
\item
If $Y$ is a riskfree asset ($\Yup=\Ydn$) then we are back with (\ref{eq:ch5.probs}).
\item
The suffix $_Y$, which we are using because we have divided values through by $Y$. Rather than finding values in dollars, we have found values as if we were paying for things in units of $Y$.
The technical term for this is \emph{change of numeraire}.
\item
This only works because the payoff is of the form: $V/Y = $ function of $(X/Y)$. Otherwise the replication weights $a_X$ and $a_Y$ are undefined.
\item
The expectation $\wex^Y$ relates to $\wex$ by
\[
\wex^Y [ V ] = \frac{\wex[Y V]}{\wex[Y]}.
\]
\end{itemize}

\subsection{Continuum limit of binomial model}

First, let us work out how to make a binomial tree work in the limit of many steps ($n\to\infty$). To make the variance of $X_T$ behave properly (neither collapse to zero, nor explode to infinity) we need to make
\begin{equation}
\frac{\mathfrak{u}+\mathfrak{d}}{2} \cdot n \mbox{ bounded }, \qquad
(\mathfrak{u}-\mathfrak{d})  \cdot \sqrt{n} \to \mbox{const} = \hsigma\sqrt{T} >0 .
\end{equation}

Now let us turn to the valuation of a contingent claim.
To understand the behaviour in the limit of many steps, we seek the price of one particular contingent claim, namely one that pays off the following at time $T$:
\[
Y_T (X_T/Y_T)^\lambda
\]
where $\lambda$ is a fixed number. Thus if $\lambda=0$, the payoff is $Y_T$, and if $\lambda=1$ it is $X_T$.
The value of such a claim is
\[
\wex \left[ \frac{Y_T}{\mma_T} (X_T/Y_T)^\lambda \right]
=
Y_0 \, \wex_Y \! \left[ (X_T/Y_T)^\lambda \right]
\]
and now we use the Binomial distribution to evaluate the expectation. At node $j$ on the tree the value of $X_T/Y_T$ is $(X_0/Y_0) \mathfrak{u}^j \mathfrak{d}^{n-j}$. So the expectation (without the prefactor of $Y_0$) is
\[
\sum_{j=0}^n {n \choose j} \phat_Y^j (1-\phat_Y)^{n-j} (\mathfrak{u}^j \mathfrak{d}^{n-j})^\lambda = \big(\phat_Y \mathfrak{u}^\lambda + (1-\phat_Y)\mathfrak{d}^\lambda\big)^n
\]
which it is convenient to rewrite as
\[
\left( \frac{\mathfrak{u}^\lambda+\mathfrak{d}^\lambda}{2} + (\phat_Y-\shalf) (\mathfrak{u}^\lambda-\mathfrak{d}^\lambda)\right)^n.
\]

Now let us set up the tree in such a way that
\[
 \left. \begin{array}{r}\mathfrak{u}\\ \mathfrak{d} \end{array} \right\} = 1 \pm \hsigma \sqrt{T/n},
 \mbox{ so that } \phat_Y=\shalf. 
\]
Using (\ref{eq:ch5.probs2}), we have by the Binomial expansion
\[
 \left. \begin{array}{r}\mathfrak{u}^\lambda \\ \mathfrak{d}^\lambda  \end{array} \right\} = 1 \pm \lambda \hsigma \sqrt{\frac{T}{n}} + \frac{\lambda(\lambda-1)}{2} \frac{\hsigma ^2T^2}{n} + \cdots
\]
So
\[
\frac{\mathfrak{u}^\lambda+\mathfrak{d}^\lambda}{2} = 1 + \frac{\lambda(\lambda-1)}{2} \frac{\hsigma ^2T}{n} + \cdots
\]
and using the limit
\[
\lim_{z\to\infty} (1+z/n)^n = e^z
\]
we deduce that in the continuum limit,
\begin{equation}
\wex^Y_0 \big[ (X_T/Y_T)^\lambda \big] = (X_0/Y_0)^\lambda e^{\lambda(\lambda-1)\hsigma^2T/2} .
\label{eq:ch5.ex_y}
\end{equation}
Now we employ a useful result\footnote{Uniqueness of inverse of the characteristic or moment-generating function. This is a standard result of complex/Fourier analysis.} from probability concerning the Normal distribution of mean $m$ and variance $v$:
\begin{equation}
Z\sim \textrm{N}(m,v) \Longleftrightarrow \ex[e^{\lambda Z}] = e^{m \lambda + v \lambda^2 /2} \mbox{ for all } \lambda.
\end{equation}
By comparing these two results (matching the coefficients of $\lambda$ and $\lambda^2$), we identify $m$ and $v$ and deduce
\[
\boxed{\mbox{Under } \wex^Y : \quad \ln(X_T/Y_T) \sim \textrm{N}\big(\ln(X_0/Y_0)-\shalf\hsigma^2T, \hsigma^2T\big).}
\]

Now as it happens we would also like to find the distribution of $\ln(X_T/Y_T)$ using yet another measure, $\wex^X$ defined analogously to $\wex^Y$:
\[
\wex^X [ V] = \frac{\wex[X V]}{\wex[X]}
\]
We have
\begin{equation}
\wex^X_0 \big[ (X_T/Y_T)^\lambda \big] = (X_0/Y_0)^\lambda e^{\lambda(\lambda+1)\hsigma^2T/2} .
\label{eq:ch5.ex_x}
\end{equation}
This can ({\sl exercise}) be obtained from (\ref{eq:ch5.ex_y}) in either of two ways: (i) replace $\lambda$ by $\lambda+1$; (ii) switch $X$ and $Y$ over and flip the sign of $\lambda$.
Consequently,
\[
\boxed{\mbox{Under } \wex^X : \quad \ln(X_T/Y_T) \sim \textrm{N}\big(\ln(X_0/Y_0)+\shalf\hsigma^2T, \hsigma^2T\big).}
\]

\subsection{Black, Scholes and Margrabe}

We have done all we need to price the `exchange option'. The payoff is
\[
\max (X_T-Y_T,0) = X_T \indic{X_T>Y_T} - Y_T \indic{X_T>Y_T}.
\]
The first term pays `$X$ or nothing' and is the expectation of $\indic{X_T>Y_T}$ under  $\wex^X$, and the second pays `$Y$ or nothing' and is its expectation under $\wex^Y$. The condition in the indicator is tantamount to $\ln(X_T/Y_T)>0$ and we have established that $\ln(X_T/Y_T)$ is Normal under both $\wex^X$ and $\wex^Y$.
Accordingly:
\begin{thm}
If $X,Y$ are tradable assets that do not pay dividends or coupons, then under the Black-Scholes assumptions the value of an option of maturity $T$ is
\begin{equation}
\boxed{\wex_0 \left[ \max \left(\frac{X_{T}-Y_{T}}{\mma_{T}},0 \right) \right] =
X_0 \Phi(d_+) - Y_0 \Phi(d_-)}
\label{eq:ch5.bs0}
\end{equation}
where
\[
\boxed{d_\pm = \frac{\ln(X_0/Y_0) \pm \half \hsigma^2 T}{\hsigma \sqrt{T}}}
\]
and $\hsigma$ is the lognormal volatility of $(X/Y)$
and $\Phi$ is the cumulative Normal distribution function, i.e.\ $\Phi(z)=\pr(Z<z)$ where $Z$ is distributed as $\mathrm{N}(0,1)$. $\Box$
\end{thm}

Verifying the put-call parity formula is easy:
\begin{eqnarray*}
C &=& X_0 \Phi(d_+) - Y_0 \Phi(d_-) \\
P &=& Y_0 \Phi(-d_-) - X_0 \Phi(-d_+)
\end{eqnarray*}
and so
\[
C - P = X_0 \big( \Phi(d_+) + \Phi(-d_+) \big) - Y_0 \big( \Phi(d_-) + \Phi(-d_-)\big) = X_0-Y_0 
\]
using the symmetry of the Normal distribution (if $Z$ is distributed $\mathrm{N}(0,1)$ then so too is $-Z$):
\[
\Phi(z)=\pr(Z<z)=\pr(-Z>-z)=1-\pr(-Z<-z)=1-\Phi(-z).
\]

We have {\bf not} made any distributional assumptions about $X_t$ and $Y_t$ individually: rather, the assumption is that $X_t/Y_t$ follows a geometric Brownian motion. 

The Black--Scholes equation is a special case where $Y$ is cash, i.e.\ $Y_0=Ke^{-rT}$.

\subsection{Deltas revisited}

When differentiating the (call) option price with respect to $X$ a `naughty' route is to say: the coefficient of $X_0$ is just $\Phi(d_+)$, so that must be the delta. Of course, this is inadmissible because it ignores that $d_\pm$ both depend on $X_0$, and two extra terms need to be calculated. But in fact these two terms cancel, so in fact $\Phi(d_+)$ is the correct answer. But why?

One route is simply to do the algebra, but that makes the whole thing look like a coincidence: we suggest, by contrast, that there is a deeper reason, not specific to the Black--Scholes setup (i.e.\ works for distributions other than lognormal). 

The reason for this result is that the two measures $\ex^X$ and $\ex^Y$ are connected in the same way that causes the cancellstion of the extra terms. Specifically, we are to show that
\begin{equation}
X_0 \pderiv{}{X_0} \pr^X[X_T>Y_T] = Y_0 \pderiv{}{X_0} \pr^X[X_T>Y_T] 
\label{eq:sens2}
\end{equation}
in a way that does not rely on the probabilities being expressed in terms of $\Phi$.
Theoretically, it should be possible to see this immediately from the definitions of the measures $\pr^X$ and $\pr^Y$, but an intermediate step seems to be beneficial, as follows; predictably, perhaps, it uses the moment-generating function.

We have, essentially from the inverse Laplace or Mellin transform,
\[
\pr^X[X_T/Y_T > 1] = \frac{1}{2\pi\I} \int \left( \frac{X_0}{Y_0} \right)^\lambda M^X(\lambda) \, \frac{d\lambda}{\lambda}
\]
with $M^X$ defined by
\[
\wex^X[(X_T/Y_T)^\lambda] = (X_0/Y_0)^\lambda M^X(\lambda) 
\]
and similarly for $M^Y$. The contour runs up the imaginary axis, avoiding the singularity at the origin by passing it on the right.
Now
\begin{eqnarray}
X_0 \pderiv{}{X_0} \wpr^X[X_T>Y_T] &=& \frac{1}{2\pi\I} \int \left( \frac{X_0}{Y_0} \right)^{\lambda} M^X(\lambda) \, d\lambda \label{eq:sensx} \\
Y_0 \pderiv{}{X_0} \wpr^Y[X_T>Y_T] &=& \frac{1}{2\pi\I} \int \left( \frac{X_0}{Y_0} \right)^{\lambda-1} M^Y(\lambda) \, d\lambda \label{eq:sensy}
\end{eqnarray}
But the relation between the two measures $\wpr^X$ and $\wpr^Y$ is 
\[
M^X(\lambda) = M^Y(\lambda+1),
\]
which follows directly from two identical expressions for the following:
\begin{eqnarray*}
\wex_0 \left[ \frac{X_T}{\mma_T} \left( \frac{X_T}{Y_T} \right)^\lambda \right] &=&
\frac{X_0^{\lambda+1}}{Y_0^\lambda} M^X(\lambda)
\\
\wex_0 \left[ \frac{X_T}{\mma_T} \left( \frac{X_T}{Y_T} \right)^\lambda \right] &=&
\wex_0 \left[ \frac{Y_T}{\mma_T} \left( \frac{X_T}{Y_T} \right)^{\lambda+1} \right] =
\frac{X_0^{\lambda+1}}{Y_0^\lambda} M^Y(\lambda+1)
\end{eqnarray*}
and so the expressions in (\ref{eq:sensx},\ref{eq:sensy}) are identical, proving (\ref{eq:sens2}).
Note that we did not need to use the fact that $M$ is an exponential-quadratic, so the argument applies more generally.


\bibliographystyle{plain}
\bibliography{}

\end{document}